\documentclass[intlimits,twoside,a4paper]{article}

\usepackage{amsmath,amssymb}
\usepackage{graphicx}
\usepackage{wrapfig}
\usepackage{siunitx}

\usepackage[T2A]{fontenc}
\usepackage[cp1251]{inputenc}

\usepackage[eqsecnum]{cmpj2}
\articletype{Proceedings Paper}

\issue{2013}{16}{3}{31703}
\doinumber{10.5488/CMP.16.31703}

\title[Electrophysical properties of PMN-PT-PS-PFN: Li ceramics]%
{Electrophysical properties of PMN-PT-PS-PFN:Li ceramics}

\author[R. Skulski \textsl{et al.}]
{R. Skulski\refaddr{label1}, D. Bochenek\refaddr{label1}, P. Niemiec\refaddr{label1}\footnote{E-mail: niemiec.przemek@gmail.com} \ ,
P. Wawrza\l{}a\refaddr{label1}, J. Suchanicz\refaddr{label2}}
\addresses{
\addr{label1} University of Silesia, Department of Materials Science, 2, \'Snie\.zna St., 41--200 Sosnowiec, Poland
\addr{label2} Pedagogical University, 2 Podchor\k{a}\.zych St., 30--084 Cracow, Poland}

\authorcopyright{R. Skulski, D. Bochenek, P. Niemiec, P. Wawrza\l{}a, J. Suchanicz, 2013}
\date{Received October 3, 2012, in final form January 16, 2013}

\begin{document}

\maketitle

\begin{abstract}
We present the technology of obtaining and the electrophysical properties of a multicomponent material \linebreak 0.61PMN-0.20PT-0.09PS-0.1PFN:Li (PMN-PT-PS-PFN:Li). The addition of PFN into PMN-PT decreases the temperature of final sintering which is very important during technological process (addition of Li decreases electric conductivity of PFN). Addition of PS i.e., PbSnO$_3$ (which is unstable in ceramic form) permits to shift the temperature of the maximum of dielectric permittivity. One-step method of obtaining ceramic samples from oxides and carbonates has been used. XRD, microstructure, scanning calorimetry measurements and the main dielectric, ferroelectric and electromechanical properties have been investigated for the obtained samples.

\pacs 77.84.Dy, 77.80.Dj, 77.80.Bh, 77.22Gm
\end{abstract}

\section{Introduction}

Pb(Mg$_{1/3}$Nb$_{2/3}$)O$_3$ (PMN) is a classic relaxor. Polarization of PMN gradually decreases with an increasing temperature in a very wide temperature range. The maximum of dielectric permittivity vs. temperature is diffused and the temperature $T_\mathrm{m}$ depends on the frequency of measurements. Macroscopic structural investigations do not exhibit the presence of a phase transition. It is widely believed that such properties of PMN are related to the existence of polar regions instead of normal ferroelectric domains.

In solid solutions ($1-x$)Pb(Mg$_{1/3}$Nb$_{2/3}$)O$_3$-$x$PbTiO$_3$ (PMN-PT) $T_\mathrm{m}$ shifts towards higher temperatures with an increasing $x$ (from about $-3$$^\circ$C for $x=0$ up to about $227$$^\circ$C for $x=0.5$). At the same time, a continuous change of the properties from relaxor to normal ferroelectric properties takes place. For low values of $x$, the hysteresis loops of PMN-PT are very narrow, while for higher $x$ the loops become wide. Phase diagram of PMN-PT based on dielectric permittivity measurements was first presented in the work by Shrout \cite{shrout}. More recently, Noheda et al. \cite{noheda}, Singh et al. \cite{singh} and Zekria \cite{zekria} improved the phase diagram of PMN-PT basing on structural investigations. More recent investigations using synchrotron radiation made by Ye et al. \cite{ye} showed almost zero changes of elementary cell parameters, which means it is very hard to estimate the phase transition temperature. Also, some of our previous works, for instance \cite{skulski,wawrzala}, concerned PMN-PT. Our samples described in this paper are based on 0.75PMN-0.25PT in which a diffused phase transition estimated by various methods takes place at temperature range from about $50$$^\circ$C to about $100$$^\circ$C.

The nature of phase transition in a solid solution ($1-x$)Pb(Fe$_{1/2}$Nb$_{1/2}$)O$_3$-$x$PbTiO$_3$ (i.e., PFN-PT) with an increasing $x$ changes from relaxor ferroelectric ($T_\mathrm{m}$ at about $127$$^\circ$C) to normal ferroelectric \cite{sunder}. At the same time, the crystal structure changes from rhombohedral to the tetragonal \cite{sunder}. The addition of PFN into PMN-PT decreases the temperature of final sintering, which is very important for lead containing materials \cite{fu}. High electric conductivity of PFN \cite{bochenek} can be decreased by addition of Li \cite{wojcik,bochenek2,bochenek3,xia}. The addition of PS i.e., PbSnO$_3$ into PMN-PT (which is unstable in ceramic form) permits to shift the temperature of the maximum of dielectric permittivity.

The main aim of this paper is to obtain and to investigate the main properties and phase transitions in 0.6075PMN-0.2025PT-0.09PS-0.1PLFN (abbreviation PMN-PT-PS-PFN:Li). One-step method has been used for obtaining our samples in which simple oxides and carbonates were used as starting components.

\section{Experimental}

The  below described PMN-PT-PS-PFN:Li samples were obtained using the classic ceramic technology from the oxides: PbO, MgO, Nb$_2$O$_5$, TiO$_2$, Fe$_2$O$_3$, SnO$_2$ and from the carbonate: Li$_2$CO$_3$. PbO and MgO were weighted with excess of 6.0 \%~mol. and 2.0 \%~mol., respectively. The initial components were mixed and milled in a planetary ball mill. At the next step, the obtained powders were pressed into pellets and synthesized ($T_\mathrm{synth}=850$$^\circ$C, $t_\mathrm{synth}=4$~h). Then, the pellets were crushed,  once more mixed and milled, and finally pressed into discs with a diameter of about 10~mm and thickness of 1~mm. The obtained discs were sintered at $T_\mathrm{s}=1050$$^\circ$C, $t_\mathrm{s}=3$~h. The final steps of the process were as follows: grinding, polishing, removing the mechanical stresses by heating, and setting silver paste electrodes.

Investigations of a crystallographic structure of the obtained ceramic samples were performed using the Philips X'Pert diffractometer. Dielectric measurements were performed during the heating (with the heating rate of about $0.5$$^\circ$C/min) using a QuadTech 1920 Precision LCR meter (frequencies from $f=0.1$~Hz to 1000~kHz). $P-E$ hysteresis loops were investigated using a virtual Sawyer-Tower bridge and Matsusada Inc. HEOPS-5B6 precision high voltage amplifier. The data were stored on a computer disc using A/D transducer card. Electromechanical measurements were carried out using D-64 Philtec Inc. optical displacement meter and high voltage amplifier (see above).

Specific heat measurements were made using a Netzsch DSC F3 Maia scanning calorimeter of the temperature range from $-150$$^\circ$C to $400$$^\circ$C under argon atmosphere at a flow rate of 30~ml/min. The specimen consisted of a single piece of ceramics of the average mass of 20~mg and was placed in an alumina crucible. The data were collected during the heating and cooling processes with constant rate of~$10$$^\circ$C/min.

\section{Results}

\begin{figure}[!h]
\centerline{
\includegraphics[width=0.8\textwidth]{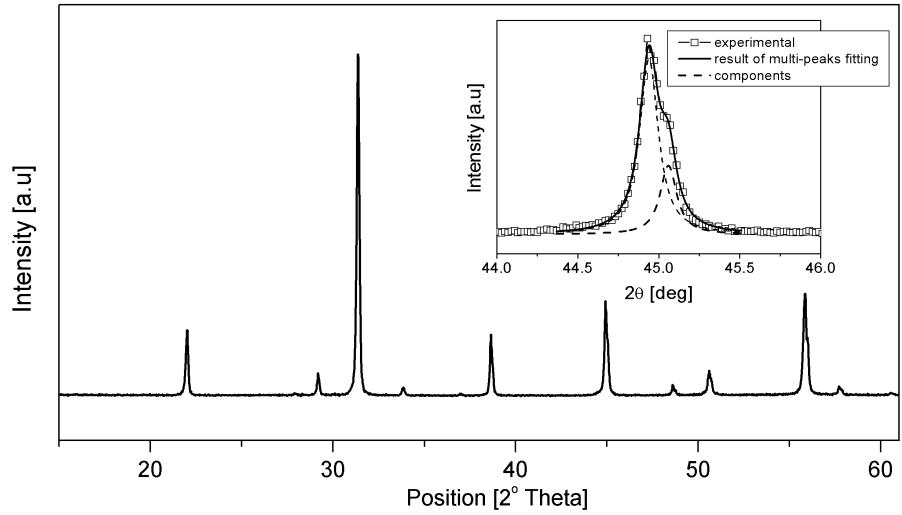}
}
\caption{XRD pattern for PMN-PT-PS-PFN:Li ceramics. In the insert~--- line (200).}\label{fig1}
\end{figure}

The XRD pattern of the obtained ceramics is presented in figure~\ref{fig1}. The result of a multicomponent analysis of the 200 maximum is presented in the insert in figure~\ref{fig1}. The 200 maximum consists of two components. However, they are very close to one another. Hence, we can assert that the crystalline structure is pseudocubic. For the angels of about 29$^\circ$, 32$^\circ$ and 49$^\circ$, the small maxima from unwanted phases are visible. This is probably a consequence of incomplete reaction during the annealing.

\begin{figure}[!t]
\centering
\includegraphics[width=0.48\textwidth]{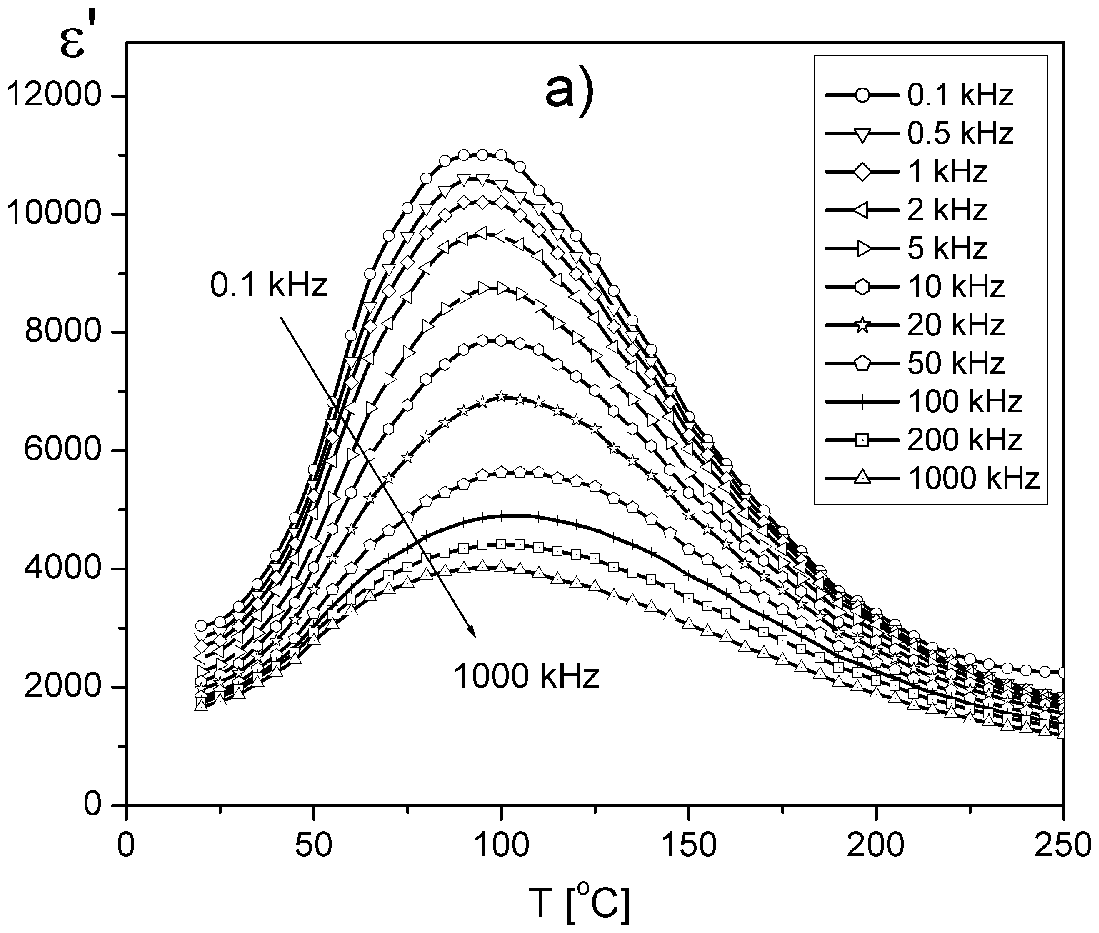}
\hspace{2mm}
\includegraphics[width=0.48\textwidth]{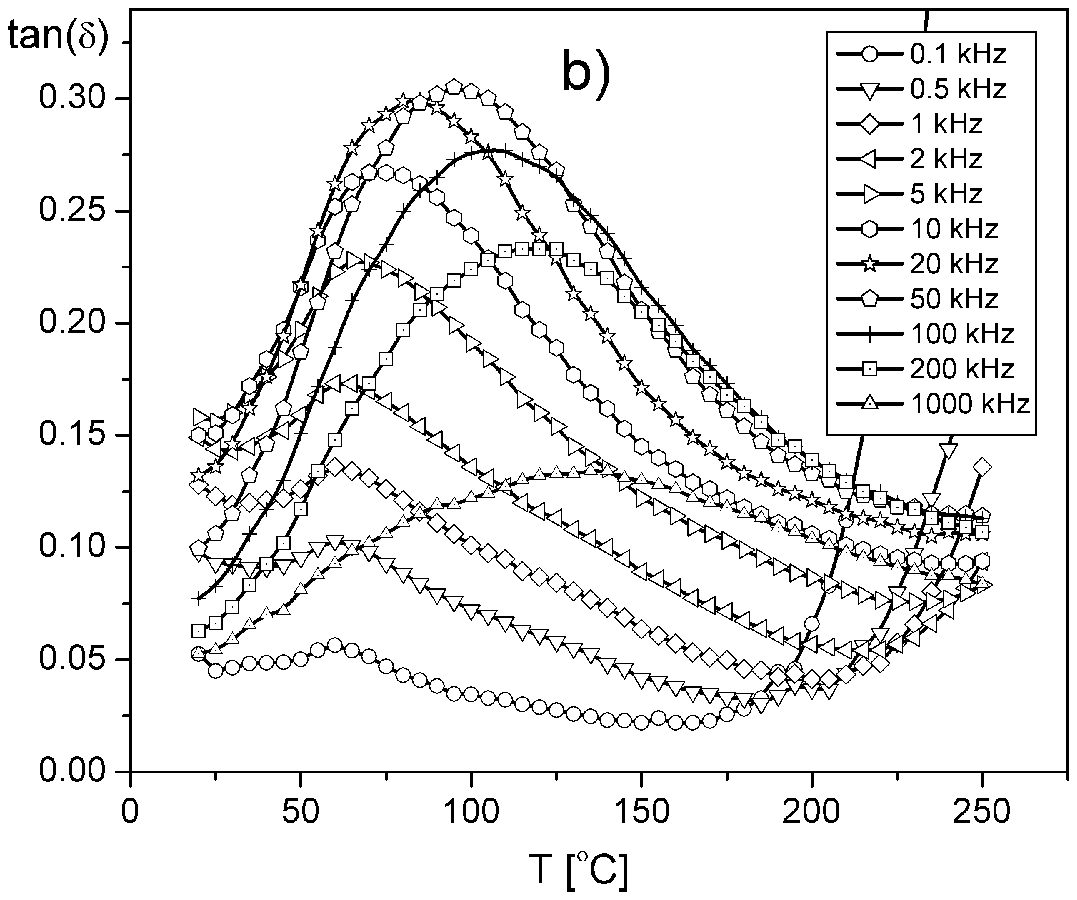}
\caption{Dependencies $\varepsilon'(T)$ (a), and tan$\delta(T)$ (b) (on heating 0.5$^\circ$C/min) for PMN-PT-PS-PFN:Li ceramics.}\label{fig2}
\end{figure}

The results of investigations of the dependencies of dielectric permittivity and dielectric loss on the temperature are presented in figure~\ref{fig2}~(a) and figure~\ref{fig2}~(b), respectively.


\begin{figure}[!h]
\centering
\includegraphics[width=0.48\textwidth]{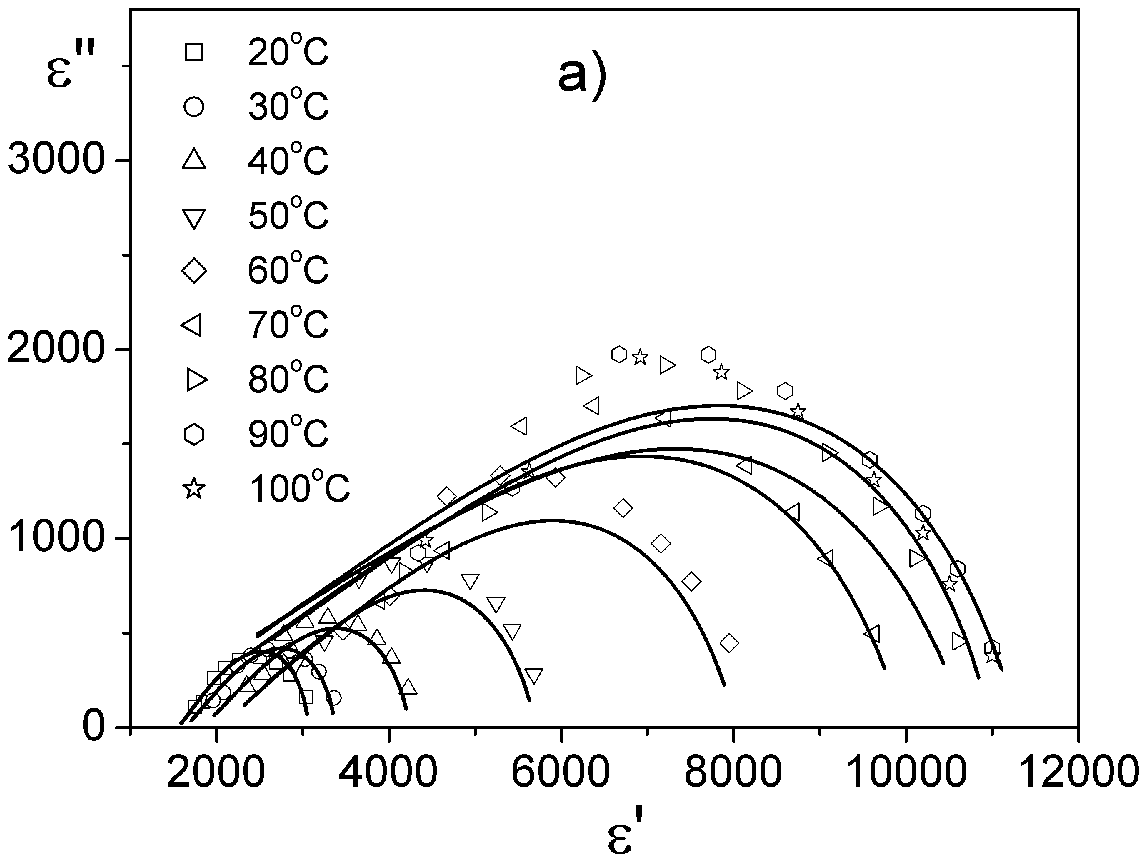}
\hspace{2mm}
\includegraphics[width=0.48\textwidth]{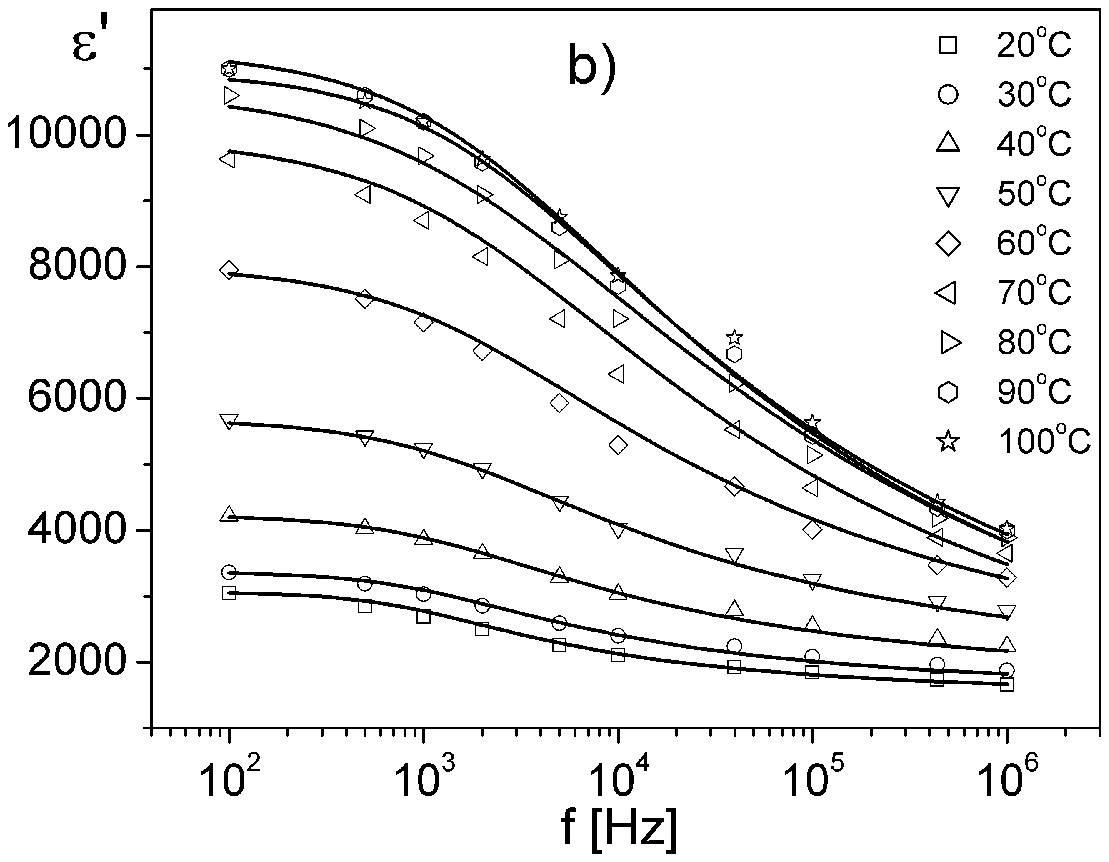}
\caption{(a) Dependencies $\varepsilon''$$(\varepsilon ')$ obtained based on the data presented in figure~\ref{fig2}. (b) Dependencies $\varepsilon'(f)$ obtained based on the data presented in figure~\ref{fig2}.}\label{fig3}
\end{figure}

The maximum of dielectric permittivity is diffused, which is typical of relaxor materials. The dispersion of dielectric permittivity is rather strong, but a shift of the maximum with frequency is not observed. At the room temperature, the values of losses of the obtained PMN-PT-PS-PFN:Li ceramics are low and increase with temperature and frequency.  Dependencies $\varepsilon''$$(\varepsilon ')$ and $\varepsilon'(f)$ are presented in figure~\ref{fig3}.

Curves from figure~\ref{fig3}~(b) have been fitted to a real part of the complex permittivity obtained from Havriliak-Negami equation \cite{havriliak}:

\begin{equation}
\label{Havriliak-Negami}
\varepsilon^* = \varepsilon_\infty + \frac{\Delta\varepsilon}{(1+\ri\omega\tau^\alpha)^\beta}\,.
\end{equation}

The results obtained in such a way are presented in figure~\ref{fig3}~(b) as solid lines. Parameters of fitting are presented in table~\ref{tab1}.

\begin{table}[!t]
\caption{Parameters of fitting the experimental data to equation (\ref{Havriliak-Negami}).}
\label{tab1}
%\vspace{2ex}
\begin{center}
\begin{tabular}{|c|c|c|c|c|c|}
\hline $T$ [$^\circ$C] & $\varepsilon_\infty$ & $\varepsilon_\mathrm{s}$ & 1/$f_0$ [s] & $\alpha$ & $\beta$ \\
\hline\hline
20 & 1553 & 3064 & 0.00108 & 0.91 & 0.38 \\

30 & 1640 & 3376 & 0.00104 & 0.93 & 0.30 \\

40 & 1837 & 4239 & 0.00079 & 0.80 & 0.35 \\

50 & 2038 & 5701 & 0.00069 & 0.75 & 0.34 \\

60 & 1950 & 8027 & 0.00061 & 0.70 & 0.32 \\

70 & 1200 & 9970 & 0.00056 & 0.66 & 0.31 \\

80 & 800 & 10700 & 0.00051 & 0.62 & 0.30 \\

90 & 1000 & 11300 & 0.00048 & 0.68 & 0.29 \\

100 & 1200 & 11000 & 0.00045 & 0.70 & 0.29 \\
\hline
\end{tabular}
\end{center}
\end{table}

\begin{figure}[h]
\centerline{
\includegraphics[width=0.5\textwidth]{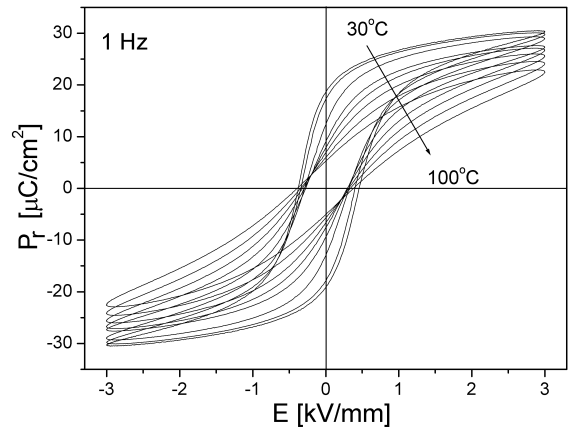}
}
\caption{$P-E$ hysteresis loops for PMN-PT-PS-PFN:Li ceramics at different temperatures (frequency 1.0~Hz).}\label{fig4}
\end{figure}

$P-E$ hysteresis loops are presented in figure~\ref{fig4}.

\begin{figure}[!h]
\centering
\includegraphics[width=0.45\textwidth]{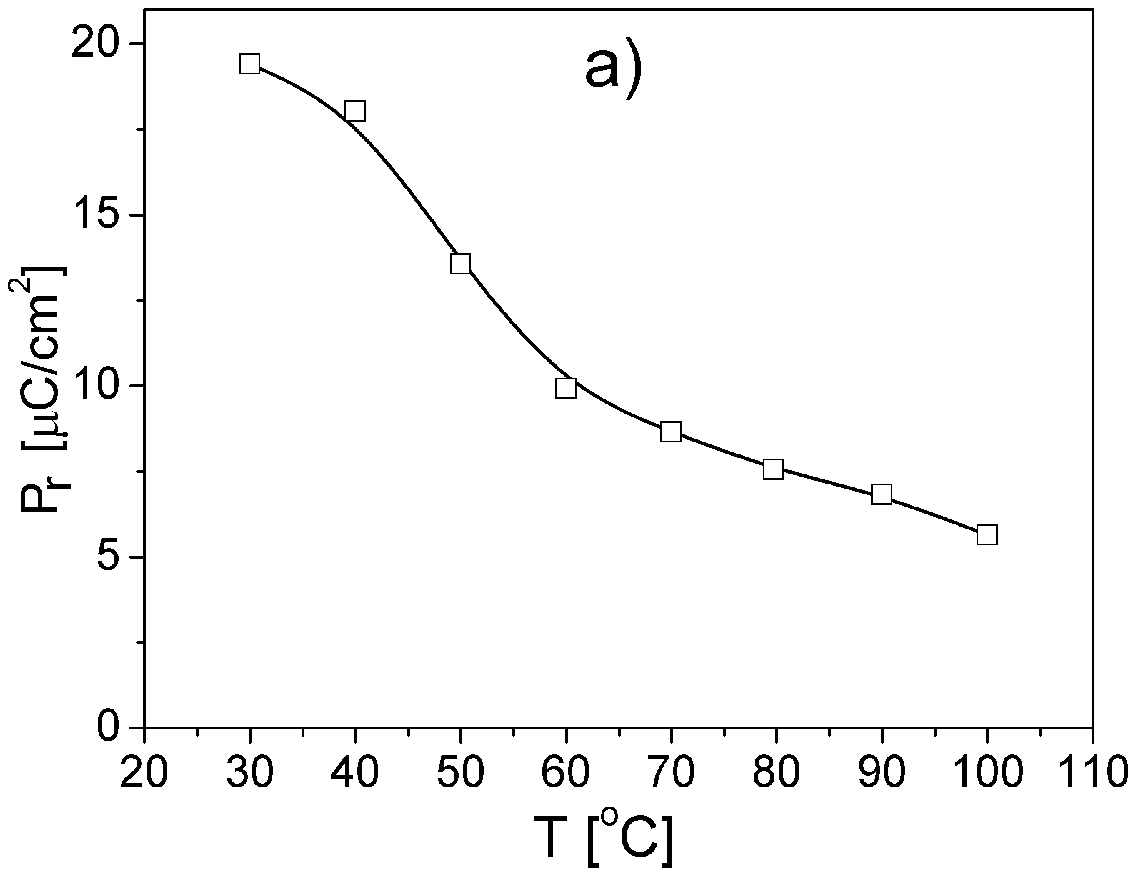}
\hspace{2mm}
\includegraphics[width=0.45\textwidth]{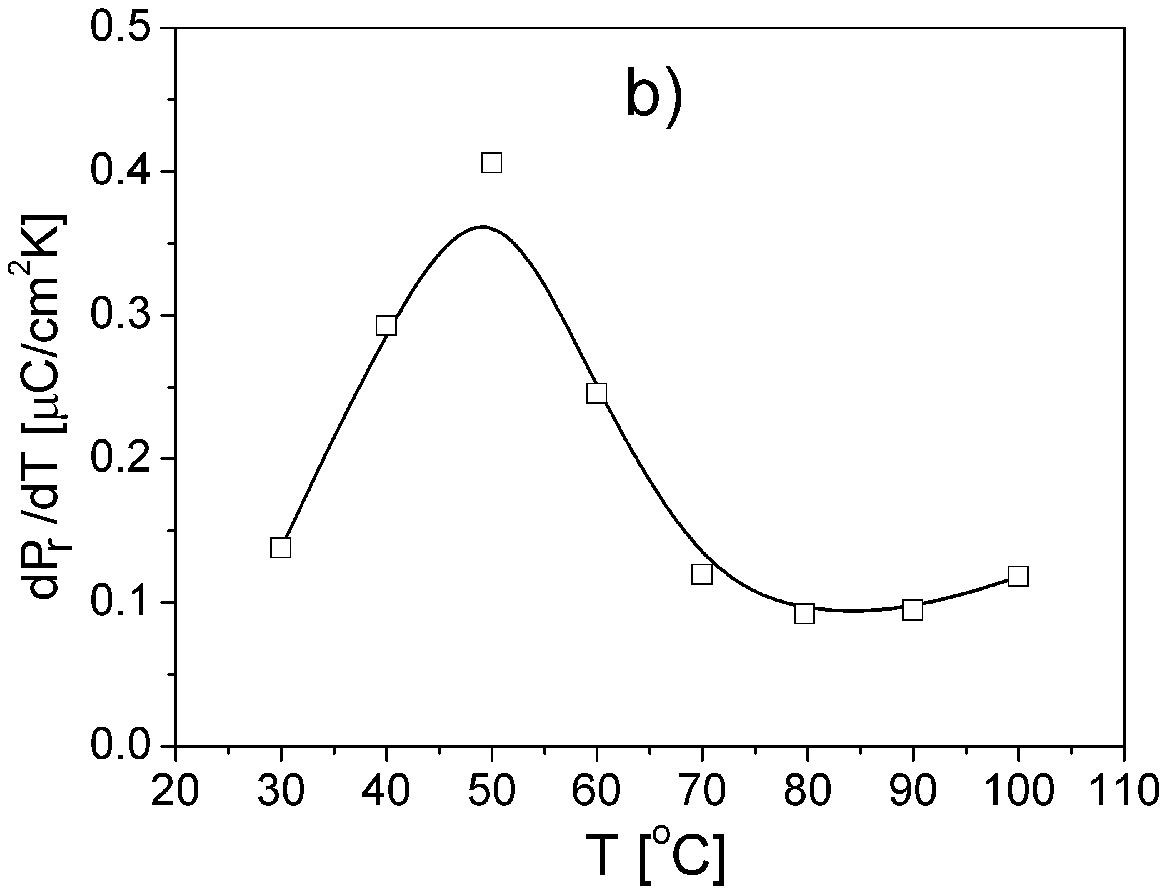}
\caption{(a) Dependency $P_\mathrm{r}(T)$ and  (b) dependency ${\partial P_\mathrm{r}(T)}/{\partial T}$ for PMN-PT-PS-PFN:Li ceramics.}\label{fig5}
\end{figure}

With an increasing temperature, the value of spontaneous polarization $P_\mathrm{s}$ decreases from about 26~\SI{}{\micro\coulomb}/cm$^2$ for 30$^\circ$C to about 17~\SI{}{\micro\coulomb}/cm$^2$ for 100$^\circ$C. The residual polarization $P_\mathrm{r}$ decreases from about 19~\SI{}{\micro\coulomb}/cm$^2$ for 30$^\circ$C to about 6~\SI{}{\micro\coulomb}/cm$^2$ for 100$^\circ$C. Coercive field $E_\mathrm{C}$ decreases from 0.4~kV/mm for 30$^\circ$C to 0.30~kV/mm for 50$^\circ$C. The increase of $E_\mathrm{C}$ to about 0.4~kV/mm (for 100$^\circ$C) is probably related to the increase of electric conductivity. The dependency $P_\mathrm{r}(T)$ is presented in figure~\ref{fig5}~(a). Figure~\ref{fig5}~(b) presents the derivative of $P$ on temperature calculated from this dependency.

The maximum of the derivative of $P_\mathrm{r}$ is observed at the temperature of about 50$^\circ$C, i.e., lower than the maximum of dielectric permittivity measured using RCL meter. A similar situation was observed, for example, for a PMN single crystal in the work \cite{fu}.

\begin{figure}[!h]
\centerline{
\includegraphics[width=0.47\textwidth]{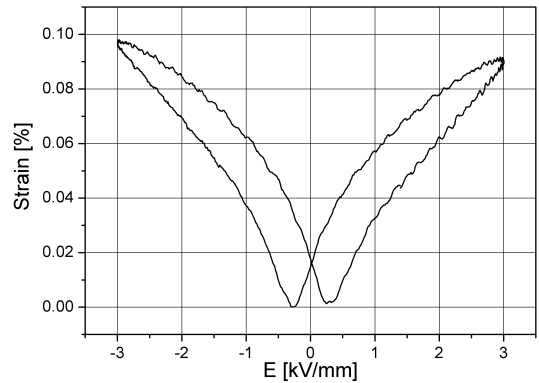}
}
\caption{Mechanical strain vs. electric field for PMN-PT-PS-PFN:Li ceramics.}\label{fig6}
\end{figure}

Figure~\ref{fig6} shows the strain loop in the function of electric field for  PMN-PT-SP-PFN:Li ceramics. The value of the $d_{33}$ coefficient measured at room temperature for the unipolar deformation using a field of about 0.5~kV/mm is equal to $355\cdot10^{-12}$~m/V and for the unipolar deformation using a field about 1~kV/mm is equal to $440\cdot10^{-12}$~m/V. Maximum value of $d_{33}$ for bipolar polarization is about $700\cdot10^{-12}$~m/V at the field of 1~kV/mm.

The results of investigations using a scanning calorimeter are presented in figure~\ref{fig6}.

\begin{figure}[!b]
\centerline{
\includegraphics[width=0.50\textwidth]{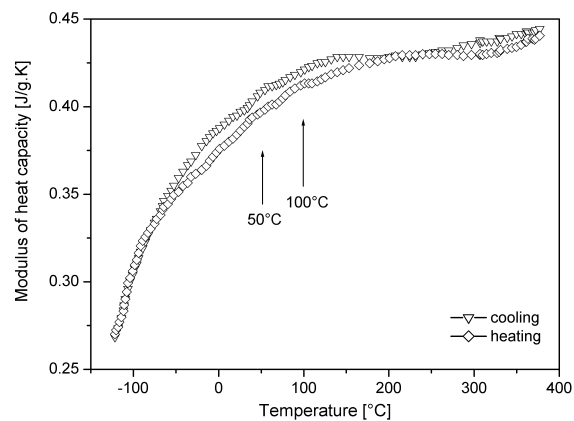}
}
\caption{The results of investigations of PMN-PT-PS-PFN:Li ceramics using a scanning calorimeter.}\label{fig7}
\end{figure}

Calorimetric investigations confirmed that the phase transition in the investigated materials is diffused. Based on the curves from figure~\ref{fig7} we observe the anomalies (indicated by arrows) suggesting phase transitions in these temperatures.

\section{Summary}

The above described   PMN-PT-PS-PFN:Li ceramics obtained by us are useful materials, for instance for applications in MLCC capacitors, although they operate in weak electric fields.
Although the obtained material possesses relaxor properties, still it  somewhat differs from the most classic relaxor of PMN. For example, at room temperature, in the XRD spectrum, a small fission of a 200 maximum can be seen, which means that at room temperature it is not 100\% pseudoregular phase (probably, a small dopant of rhombohedral phase occurs). Despite the strong dependency of $\varepsilon'(T)$ on frequency and a typical Debye relaxation, practically no shift of $T_\mathrm{m}$ with frequency takes place. The application of a constant electric field (i.e., polarization) changes the dependencies of $\varepsilon'(T)$. This is probably due to the fact that at room temperature a normal ferroelectric domain structure exists. This is also confirmed by a relatively narrow hysteresis loop typical of soft ferroelectrics which become less saturated at higher temperatures. Temperature of the phase transition calculated from the hysteresis loop is about 50$^\circ$C lower than $T_\mathrm{m}$ temperature. This can be explained in the following way. The ferroelectric domains become smaller with an increasing temperature and are divided into nanodomains/polar regions. The calculated value of piezoelectric coefficient is rather high $d_{33}$ which is also typical of ferroelectrics. This might be a problem if the usage of the obtained material in high voltage pulse MLCC capacitors is considered.

%\bibliographystyle{cmp}
%\bibliography{mybibdb}

\ukrainianpart

\title{Електрофізичні властивості керамік PMN-PT-PS-PFN:Li}

\author{Р. Скульскі\refaddr{label1}, Д. Бохенек\refaddr{label1}, П. Нємєц
\refaddr{label1}, П. Вавжала \refaddr{label1}, Я. Суханич\refaddr{label2}}
\addresses{
\addr{label1} Сілезький університет, відділ матеріалознавства,
Сосновєц, Польща \addr{label2} Педагогічний університет, Краків,
Польща}

 \makeukrtitle
\begin{abstract}

Ми представляємо технологію отримання і електрофізичні властивості
багатокомпонентного матеріалу  0.61PMN-0.20PT-0.09PS-0.1PFN:Li
(PMN-PT-PS-PFN:Li). Додавання PFN в  PMN--PT понижує температуру
кінцевого спікання, яка є дуже важливою підчас технологічного
процесу (додавання Li понижує електричну провідність PFN). Додавання
PS а саме, PbSnO$_3$ (який є нестійким в керамічному вигляді)
дозволяє зсунути температуру максимуму діелектричної
сприйнятливості. Використано однокроковий метод отримання керамічних
зразків з оксидів і карбонатів. Для отриманих зразків вивчено XRD,
мікроструктуру, проведено скануючі калометричні вимірювання і досліджено головні
діелектричні, сегнетоелектричні і електромеханічні властивості.

\keywords релаксор, сегнетоелектрики, кераміка, конденсатор,
фазовий перехід
\end{abstract}
  \end{document}